\documentclass[twocolumn]{aastex63}
\usepackage{times}
\usepackage{amsmath}
\usepackage{textcomp}
\usepackage{amssymb}
\usepackage{natbib}
\usepackage{float}
\usepackage{color}
\usepackage{graphicx}
\usepackage{epstopdf}
\newcommand{\Msun}{\ensuremath{M_{\odot}}}
\newcommand{\mbh}{$M_{\rm BH}$}
\newcommand{\ld}{$L_{\rm disk}$}
\newcommand{\lum}{erg\,s$^{-1}$}

\newcommand{\nustar}{{NuSTAR}}

\newcommand{\ergflux}{\mbox{${\rm \, erg \,\, cm^{-2} \, s^{-1}}$}}
\newcommand{\gm}{$\gamma$}
\newcommand{\MgII}{Mg{\sevenrm II}}
\newcommand{\FeII}{Fe{\sevenrm II}}
\newcommand{\NIIab}{[N{\sevenrm\,II}]\,$\lambda\lambda$6549,6585}
\newcommand{\SIIab}{[S{\sevenrm\,II}]\,$\lambda\lambda$6718,6732}
\newcommand{\halpha}{H{$\alpha$}}

 \font\sevenrm=cmr7 scaled 1000

\submitjournal{ApJ}

\shorttitle{PMN J1310$-$5552}
\shortauthors{Vaidehi S. Paliya}

\begin{document}
\title{PMN~J1310$-$5552: A Gamma-ray Emitting Blazar Candidate Harboring A Cosmic Monster}

\correspondingauthor{Vaidehi S. Paliya}

\author[0000-0001-7774-5308]{Vaidehi S. Paliya}
\affiliation{Inter-University Centre for Astronomy and Astrophysics (IUCAA), SPPU Campus, 411007, Pune, India}

\email{vaidehi.s.paliya@gmail.com}

\begin{abstract}
Relativistic jets manifest some of the most intriguing activities in the nuclear regions of active galaxies. Identifying the most powerful relativistic jets permits us to probe the most luminous accretion systems and, in turn, the most massive black holes. This paper reports the identification of one such object, PMN~J1310$-$5552 ($z=1.56$), a blazar candidate of uncertain type detected with the Fermi Large Area Telescope (LAT) and Swift Burst Alert Telescope. The detection of broad emission lines in its optical spectra taken with the X-Shooter and Goodman spectrographs classifies it to be a flat-spectrum radio quasar. The analysis of the Goodman optical spectrum has revealed PMN~J1310$-$5552 harbors a massive black hole (log scale \mbh$=9.90\pm0.07$, in \Msun) and luminous accretion disk (log scale \ld$=46.86\pm0.03$, in \lum). The fitting of the observed big blue bump with the standard accretion disk model resulted in the log scale  \mbh$=9.81^{+0.19}_{-0.20}$ (in \Msun) and \ld$=46.86^{+0.09}_{-0.09}$ (in \lum), respectively. These parameters suggest PMN~J1310$-$5552 hosts one of the most massive black holes and the most luminous accretion disks among the blazar population. The physical properties of this enigmatic blazar were studied by modeling the broadband spectral energy distribution considering the data from NuSTAR, Swift, Fermi-LAT, and archival observations. Overall, PMN~J1310$-$5552 is a powerful `MeV' blazar with physical parameters similar to other members of this unique class of blazars. These results provide glimpses of monsters lurking among the unknown high-energy emitters and demonstrate the importance of ongoing wide-field sky surveys to discover them.

\end{abstract}


\section{Introduction}{\label{sec:Intro}}
Active Galactic Nuclei (AGN) hosting closely aligned, i.e., beamed, relativistic jets are termed blazars. They are classified as flat spectrum radio quasars (FSRQs), and BL Lac objects based on the rest-frame equivalent width (EW) of the emission lines in their optical spectra, with the latter exhibiting weak or no lines \citep[EW$<$5\AA; e.g.,][]{1991ApJ...374..431S}. Recent studies have proposed a more physically motivated classification based on the mass accretion rate in the Eddington unit. FSRQs typically exhibit radiativelly efficient accretion with $L_{\rm acc}/L_{\rm Edd}>0.01$ \citep[][]{2011MNRAS.414.2674G}. Moreover, the mass accretion rate is found to be strongly correlated with the Compton dominance\footnote{Compton dominance is defined as the ratio of the inverse Compton to synchrotron peak luminosities.} with FSRQs having large Compton dominance \citep[$>$1;][]{2021ApJS..253...46P}. Indeed, population studies have suggested that the diverse observational features identified in the broadband spectral energy distribution (SED) of blazars can be attributed to the mass accretion rate in Eddington units \citep[cf.][]{2014Natur.515..376G}.

Blazars dominate the extragalactic \gm-ray sky observed by the Large Area Telescope (LAT) onboard Fermi Gamma-ray Space Telescope \citep[][]{2009ApJ...697.1071A}. This is likely due to the Doppler boosting of the radiation emitted from the beamed relativistic jet. Among the blazar population, both FSRQS and BL Lac objects have been detected \citep[e.g.,][]{2022ApJS..263...24A}. However, there are still $\sim$39\% \gm-ray sources whose multiwavelength properties are similar to blazars, though their classification remains uncertain due to the lack of optical spectroscopic information. Such objects are termed as blazar candidates of uncertain type or BCUs \citep[cf.][]{2011ApJ...743..171A}.

The bolometric power of blazars has been found to exhibit an anti-correlation with the peak frequency of the synchrotron and inverse Compton emission \citep[e.g.,][]{2010ApJ...710...24S}. However, this observation could be due to selection biases and, indeed, exceptions have been reported \citep[e.g.,][]{2012MNRAS.420.2899G,2020ApJ...903L...8P}. Nevertheless, focusing on the average blazar population, the low-luminous BL Lacs usually have high-energy emission peaking in the GeV-TeV energy range. On the other hand, the most powerful blazars typically have their high-energy SED peaking in the hard X-ray-to-soft \gm-ray band, i.e., at $\sim$MeV energies \citep[cf.][]{2010MNRAS.405..387G,2017ApJ...837L...5A}. Among the currently operating X-ray missions, the Burst Alert Telescope (BAT, operating range 14$-$195 keV) onboard Neil Gehrels Swift Observatory is probably the best-observing facility to detect and identify the most luminous blazar jets. Furthermore, there have been active optical spectroscopic followups of the Swift-BAT detected AGN, thus ensuring that high-quality optical spectra are available for a large sample of BAT AGN \citep[][]{2022ApJS..261....2K}. Additionally, since the jet power correlates with the accretion luminosity, the most powerful jets are expected to be associated with luminous accretion disks and, in turn, with massive black holes. Indeed, Swift-BAT detected blazars hosting powerful relativistic jets are found to be harboring black holes with mass often exceeding $10^9$ \Msun~\citep[][]{2010MNRAS.405..387G,2019ApJ...881..154P}.

Considering the facts mentioned above, the author systematically searched to identify the Swift-BAT detected Fermi-BCUs hosting massive black holes. The fourth data release of the fourth Fermi-LAT \gm-ray source catalog (4FGL-DR4) contains 1624 BCUs \citep[][]{2020ApJS..247...33A,2023arXiv230712546B}. Moreover, the second data release of the BAT AGN Spectroscopic Survey AGN catalog (BASS-DR2) provides the optical spectroscopic measurements of 858 hard-X-ray-selected AGNs from the Swift-BAT 70-month source catalog \citep[][]{2013ApJS..207...19B,2022ApJS..261....2K}. Among them, 790 sources have black hole mass ($M_{\rm BH}$) estimated following several techniques, e.g., single-epoch virial relation and host galaxy stellar velocity dispersion \citep[][]{2022ApJS..261....4O,2022ApJS..261....5M,2022ApJS..261....6K}. The 4FGL-DR4 and BASS-DR2 catalogs were cross-matched using a $5^{\prime\prime}$ search radius leading to 87 matches. Considering only Fermi-BCUs and those having $M_{\rm BH}>10^9$ \Msun, one object 4FGL J1310.7$-$5553 or SWIFT~J1310.9$-$5553 was identified. This object is associated to the radio source PMN J1310$-$5552 \citep[see, e.g.,][]{2021ApJ...916...28T}. The BASS-DR2 catalog reports its black hole mass to be $M_{\rm BH}=1.05\times10^{10}$ \Msun~\citep[][]{2022ApJS..261....5M}. The derived $M_{\rm BH}$ indicates PMN~J1310$-$5552 to be a peculiar blazar candidate worth studying. Indeed, finding a jetted AGN hosting a more than ten billion solar mass black hole is extremely rare. Therefore, multi-frequency observations of this enigmatic blazar candidate, including new Nuclear Spectroscopic Telescope Array (\nustar) and Swift datasets, were analyzed, and the results are presented in this paper. Particular attention was given to accurately derive the central engine parameters, i.e., $M_{\rm BH}$ and accretion disk luminosity ($L_{\rm disk}$). A brief introduction of PMN~J1310$-$5552 is provided in Section~\ref{sec2}. Techniques adopted to compute its central engine parameters and obtained results are described in Section~\ref{sec3}. Section~\ref{sec4} elaborates the broadband physical properties of PMN~J1310$-$5552, and overall results are summarized in Section~\ref{sec5}. Throughout, a flat cosmology with $H_0 = 70~{\rm km~s^{-1}~Mpc^{-1}}$ and $\Omega_{\rm M} = 0.3$ was adopted.

\section{PMN J1310$-$5552}\label{sec2}
\begin{figure*}
\hbox{\hspace{1.0cm}
\includegraphics[scale=0.65]{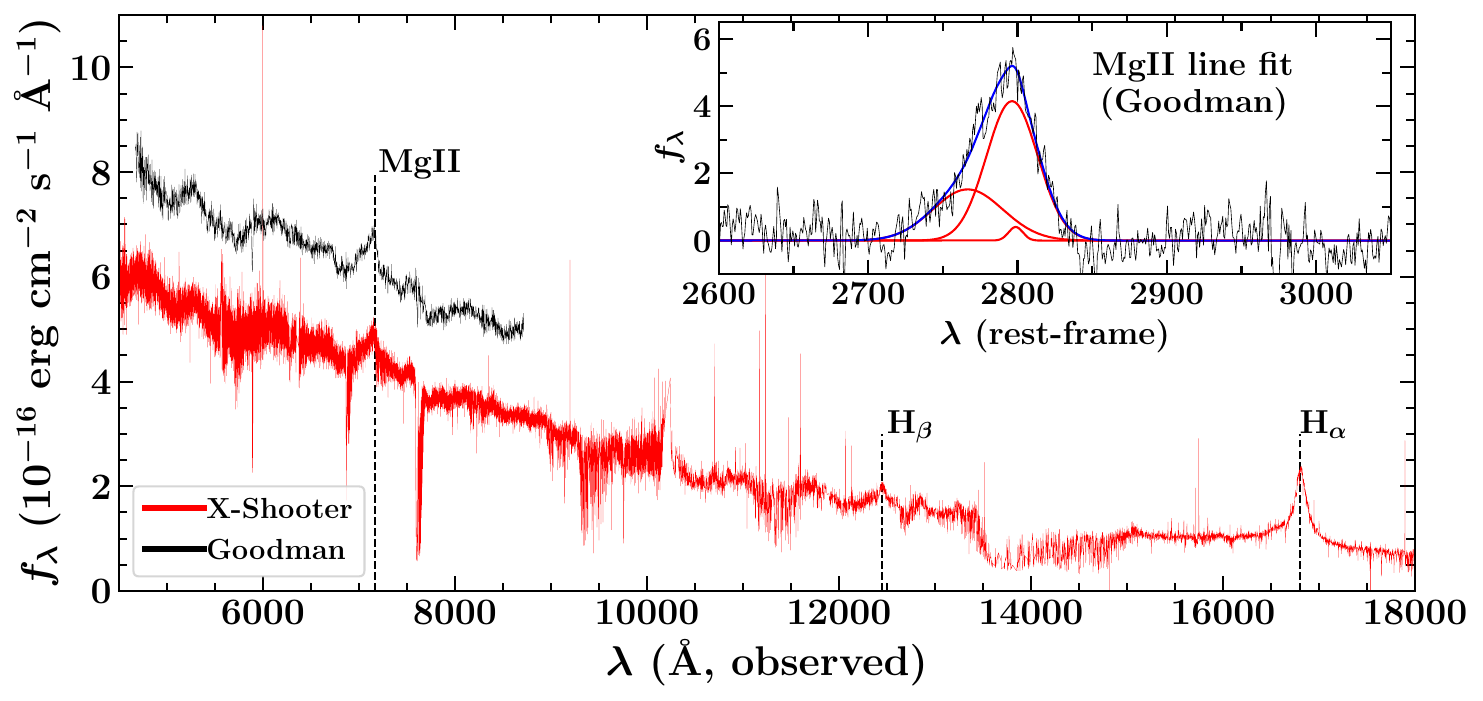}
}
\caption{The optical spectrum of PMN~J1310$-$5552 taken with X-Shooter (red) and Goodman (black) spectrographs. The inset shows the multi-Gaussian fitting done on the continuum subtracted \MgII~emission line using {\tt PyQSOFit}. The inset axes units are the same as the main panel.} \label{fig:1}
\end{figure*}

This bright, flat radio spectrum quasar was first identified in the Parkes-MIT-NRAO sky survey \citep[$F_{\rm 1.4GHz}=347\pm19$ mJy;][]{1994ApJS...90..173G,2021A&A...655A..17S}. It is located in the Galactic plane ($|b|=6^{\circ}.9$) and was found to be a hard X-ray emitter by the INTEGRAL satellite \citep[][]{2007ApJS..170..175B,2010A&A...523A..61K}. This source was also detected by the Swift-BAT \citep[$F_{\rm 14-195~keV}=2.47^{+0.31}_{-0.28}\times10^{-11}$ \ergflux, photon index $\Gamma_{\rm 14-195~keV}=1.56^{+0.31}_{-0.30}$;][]{2022ApJS..261....4O}. \citet[][]{2008A&A...482..113M} obtained an optical spectrum of PMN J1310$-$5552 with the 3.6 m telescope at the ESO-La Silla Observatory, Chile. They reported its redshift to be $z=0.104$, and its optical spectrum had stellar absorption features; however, it did not appear to be AGN-like or like their host galaxies. Given the proximity of the source to the Galactic plane, confusion is likely in identifying the correct optical counterpart. Later, \citet[][]{2022ApJS..261....6K} procured high-quality optical spectra of this object with X-Shooter and Goodman spectrographs mounted at the 8.2 m Very Large Telescope and 4.1 m Southern Astrophysical Research (SOAR) Telescope, respectively. The source redshift was estimated to be $z=1.56$, and the optical spectrum was dominated by a blue continuum and broad emission lines, typically observed from Type 1 AGN (see Figure~\ref{fig:1}). 

With a more accurate distance measurement available, the $k$-corrected 14$-$195 keV luminosity of PMN~J1310$-$5552 turns out to be (2.64$\pm$0.81)$\times10^{47}$ \lum. In the high-energy \gm-ray band (0.1$-$300 GeV), this source exhibits a steep falling spectrum (photon index $\Gamma_{\rm 0.1-300~GeV}=2.73\pm0.12$) and has an average \gm-ray luminosity of (1.12$\pm$0.32)$\times10^{47}$ \lum~\citep[][]{2020ApJS..247...33A,2023arXiv230712546B}. A more luminous hard X-ray emission, together with the observation of the flat, rising X-ray spectrum and a soft \gm-ray spectrum, indicates PMN J1310$-$5552 to be a typical `MeV' blazar \citep[][]{1995A&A...293L...1B,2017ApJ...837L...5A,2020ApJ...889..164M}.

\section{Central Engine}\label{sec3}
\subsection{Optical Spectroscopic Analysis}
The commonly adopted method to estimate the mass of the central supermassive black hole is the analysis of the single epoch optical spectrum assuming the broad line region (BLR) to be virialized \citep[see, e.g.,][and references therein]{2011ApJS..194...45S}. \citet[][]{2022ApJS..261....5M} analyzed the X-Shooter optical spectrum of PMN~J1310$-$5552. For the \MgII~emission line, which is thought to be a reliable estimator for $M_{\rm BH}$ calculation \citep[e.g.,][]{2015ApJ...807L...9W}, they reported the full width at half maximum (FWHM) of 4586$^{+135}_{-50}$ km s$^{-1}$ and the logarithmic continuum luminosity at 3000 \AA~($L_{\rm 3000}$, in \lum) of 46.90$^{+0.01}_{-0.01}$. The derived log scale $M_{\rm BH}$ was 10.02$^{+0.02}_{-0.01}$ (in \Msun). They also reported a logarithmic $M_{\rm BH}$ value of $8.03\pm0.41$ (in \Msun) based on the \halpha~emission line measurements. However, since \halpha~emission line based $M_{\rm BH}$ should be treated with caution, possibly due to blending of \NIIab~and sometime \SIIab, this was not considered in the further analysis. Moreover, \citet[][]{2022ApJS..261....2K} reported the bolometric luminosity ($L_{\rm bol}$) of PMN J1310$-$5552 to be $3.8\times10^{48}$ \lum~which gives an Eddington ratio ($L_{\rm bol}/L_{\rm Edd}$) of 281.46 using the \halpha~based \mbh~measurement. This value is larger than the largest Eddington ratio reported for quasars observed with the Sloan Digital Sky Survey \citep[][]{2022ApJS..263...42W}, hence unlikely to be correct.

\begin{figure*}
\hbox{
\includegraphics[scale=0.55]{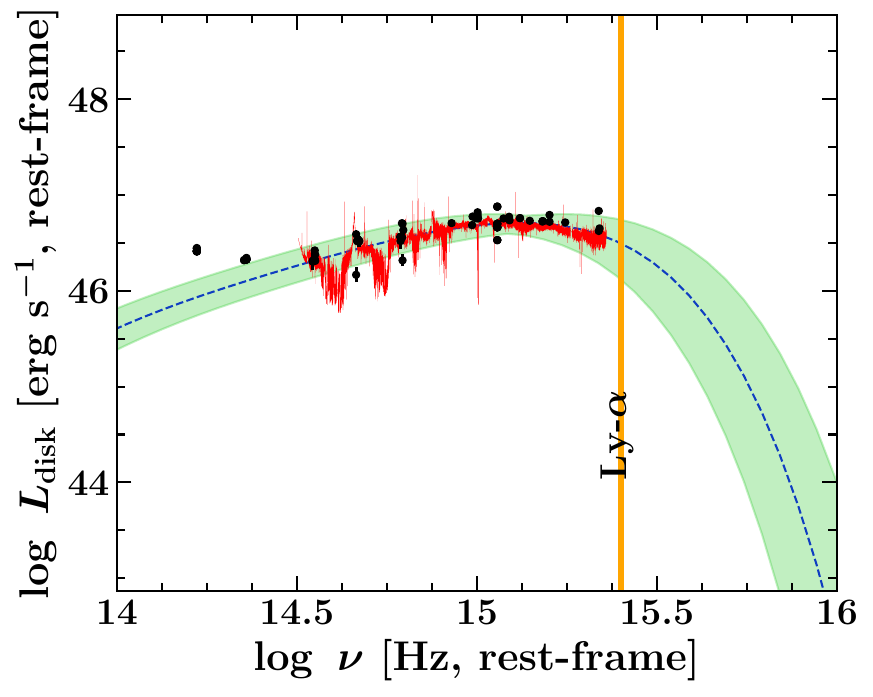}
\hspace{-0.3cm}
\includegraphics[scale=0.58]{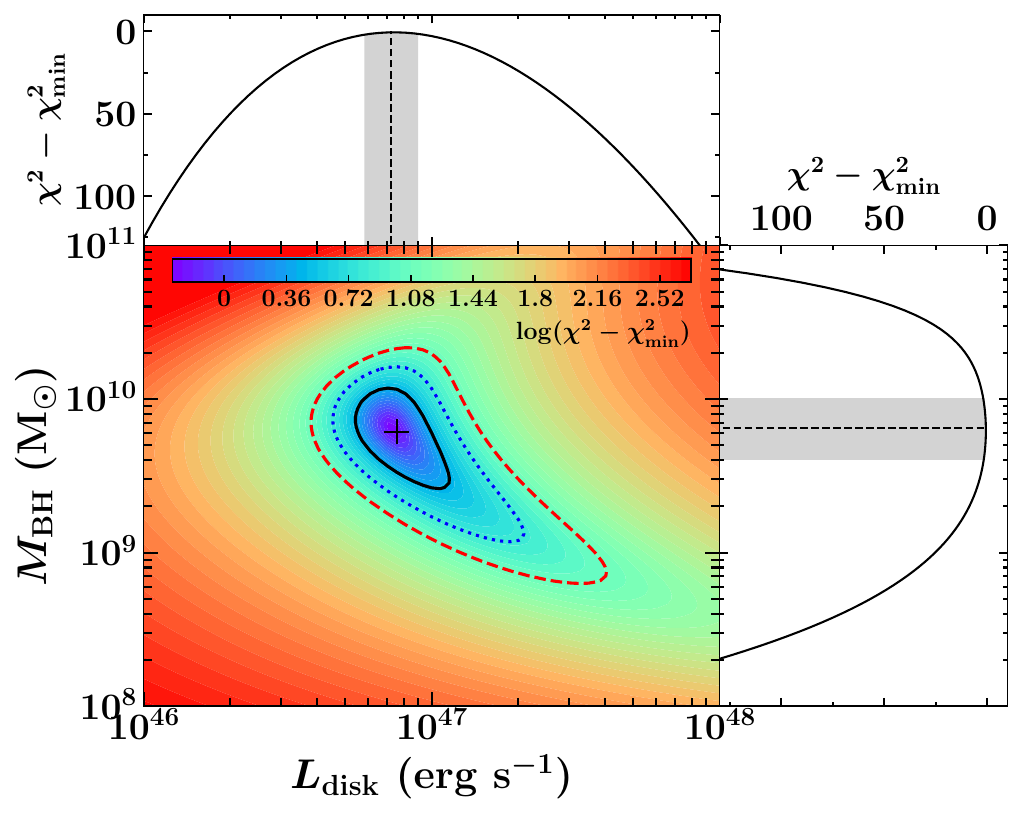}
}
\caption{Left: Optical SED of PMN~J1310$-$5552 fitted with the standard accretion disk model. The red line refers to the X-Shooter spectrum which is shown only for visualization purpose. The fitted model (blue dashed line) corresponds to the optimized \ld~and \mbh~values (`+' mark on the right panel). The green shaded area refers to the 1$\sigma$ uncertainty in the fitted model. Right: The $\chi^2$ grid of \ld~and \mbh. The black solid, blue dotted, and red dashed lines represent the confidence contours at 1$\sigma$, 2$\sigma$, and 3$\sigma$ levels, respectively. The side plots refer to $\chi^2$ variations as a function of the considered parameters. The grey shaded region shows the 1$\sigma$ uncertainty in the best-fit parameters which are highlighted with dashed lines.} \label{fig:1a}
\end{figure*}

In this work, the optical spectrum of PMN~J1310$-$5552 taken with the Goodman spectrograph was analyzed using the software {\tt PyQSOFit}\footnote{\url{https://github.com/legolason/PyQSOFit}} \citep[][]{2018ascl.soft09008G}. This tool models the emission lines and continuum emission following a $\chi^2$ or Markov Chain Monte Carlo (MCMC) based fitting technique. The salient features of the pipeline are briefly described here, and the details can be found in \citet[][]{2019ApJS..241...34S}. The spectrum was brought to the rest-frame using the source redshift and corrected for the Galactic extinction using the dust map of \citet[][]{1998ApJ...500..525S} and extinction curve from \citet[][]{1989ApJ...345..245C}. The continuum was modeled using a power-law, a third-order polynomial, and UV \FeII~template around \MgII~line \citep[][]{2001ApJS..134....1V} and subtracted from the spectrum, leaving a line-only spectrum. The \MgII~line was fitted with three Gaussian functions, two for the broad and one for the narrow components (see, inset in Figure~\ref{fig:1}). The uncertainties were estimated following an MCMC approach. 

The FWHM of the \MgII~emission line was estimated to be 4947$\pm$486 km s$^{-1}$. On the other hand, the derived logarithmic continuum luminosity at 3000 \AA~was 47.02$\pm$0.01 (in \lum). The following empirical relation was adopted to compute the virial $M_{\rm BH}$:

\begin{equation}\label{eqn:virial_estimator}
\log \left({M_{\rm BH} \over M_\odot}\right)
=A+B\log\left({\lambda L_{3000} \over 10^{44}\,{\rm
erg\,s^{-1}}}\right)+C\log\left({\rm FWHM\over km\,s^{-1}}\right) ,
\end{equation}

where $\lambda L_{3000}$ is the continuum luminosity at 3000 \AA. The coefficients $A$, $B$, and $C$ were taken as 1.816, 0.584, and 1.712, respectively, following \citet[][]{2012ApJ...753..125S}. The logarithmic \mbh~of the central supermassive black hole in PMN J1310$-$5552 was found to be 9.90$\pm$0.07 (in \Msun), which is consistent with that estimated by \citet[][]{2022ApJS..261....5M} using X-Shooter data.

The BLR luminosity can be inferred from the emission line luminosity, assuming the latter emits a certain fraction of the former. Fixing a reference value of 100 to Ly$\alpha$ emission and adding the line ratios (with respect to Ly$\alpha$) reported in \citet[][]{1991ApJ...373..465F} and \citet[][]{1997MNRAS.286..415C} gave the total BLR fraction $\langle L_{\rm BLR} \rangle  = 555.77\sim 5.6 {\rm Ly\alpha}$.  Then, the following equation can be used:

\begin{equation}
L_{\rm BLR} = L_{\rm \MgII} \times {\langle L_{\rm BLR} \rangle  \over L_{\rm rel. frac.}},
\end{equation}

where $L_{\rm \MgII}$ is the \MgII~line luminosity and $L_{\rm rel.frac.}$ is the line ratio \citep[=$34\pm22$, with respect to the Ly$\alpha$ flux;][]{1991ApJ...373..465F}. The derived logarithmic BLR luminosity is 45.86$\pm$0.03 (in \lum). Assuming the BLR to reprocess 10\% of the accretion disk radiation, the estimated \ld~is 46.86$\pm$0.03 (in \lum) or $\sim$7\% of the Eddington luminosity (1$\times$10$^{48}$ \lum). All of the quoted uncertainties were derived following the standard error propagation theory \citep[][]{1969drea.book.....B} and are purely statistical in nature. They do not take into account the possible inherent systematics associated with the virial calibration relations. For example, the virial black hole mass calculation methods typically have an uncertainty of $\sim$0.4 dex \citep[][]{2006ApJ...641..689V,2011ApJS..194...45S}.

\begin{figure*}
\hbox{\hspace{0.6cm}
\includegraphics[scale=0.4]{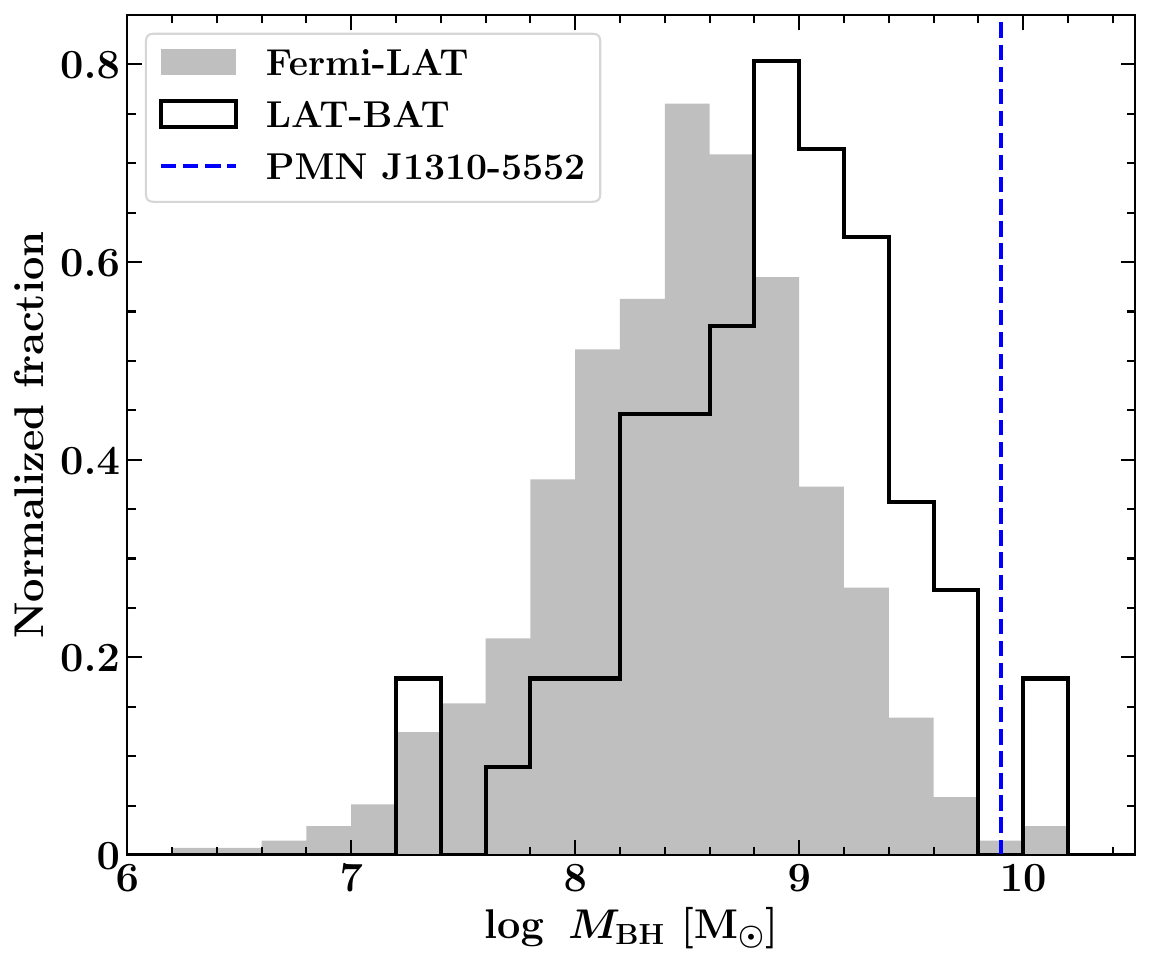}
\includegraphics[scale=0.4]{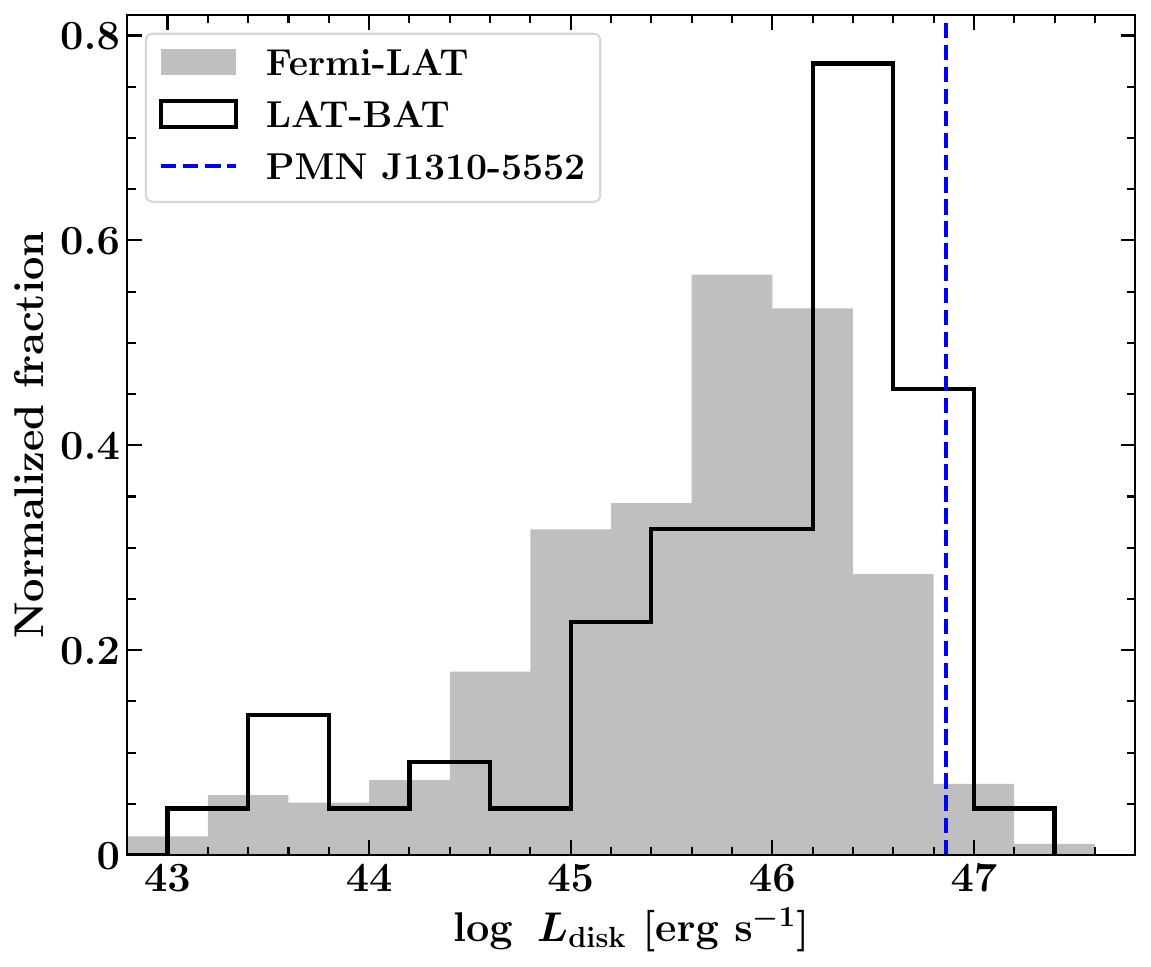}
}
\caption{The \mbh~(left) and \ld~(right) histograms for blazars detected with the Fermi-LAT and Swift-BAT (black empty) and blazars detected only with the Fermi-LAT (grey filled). The location of PMN~J1310$-$5552 is shown with the blue dashed line.} \label{fig:2}
\end{figure*}

\subsection{Modeling the Big Blue Bump}\label{subsec:bhmass}
Another technique to constrain the central engine parameters is by modeling the big blue bump peaking in the optical-ultraviolet (UV) band with a geometrically thin, optically thick accretion disk model \citep[][]{1973A&A....24..337S}. The emission profile of the accretion disk radiation can be considered as a multi-temperature blackbody parameterized as follows \citep[][]{2002apa..book.....F}

 \begin{equation}
 \label{eq:disk_flux}
 F_\nu = \nu^3\frac{4\pi h \cos \theta_{\rm v} }{c^2 {d_l}^2}\int_{R_{\rm i}}^{R_{\rm o}}\frac{R\,{\rm d}R}{e^{h\nu/kT(R)}-1}
 ,
\end{equation}
where $\theta_{\rm v}$ is the angle between the jet axis and the line of sight, $d_l$ is the luminosity distance, and $R_{\rm o}$ and $R_{\rm i}$ are the outer and inner radii of the disk, assumed as 500$R_{\rm Sch}$ and 3$R_{\rm Sch}$, respectively. The temperature profile of the disk is given as follows.

\begin{equation}
\label{eq:temp_profile}
T^{4}(R)\, =\, {  3 R_{\rm Sch}  L_{\rm disk }  \over 16 \pi\zeta_{\rm acc}\sigma_{\rm SB} R^3 }  
\left[ 1- \left( {3 R_{\rm Sch} \over  R}\right)^{1/2} \right],
\end{equation}
where $\zeta_{\rm acc}$ is the accretion efficiency, considered here as 10\%, and $\sigma_{\rm SB}$ is the Stefan-Boltzmann constant. There are only two variables in Equation~\ref{eq:disk_flux}, \mbh~(through $R_{\rm Sch}$) and \ld, under the assumption of fixed viewing angle and accretion efficiency.

Since the optical-UV emission of PMN~J1310$-$5552 reveals a prominent big blue bump, the above hypothesis can be applied to derive the \ld~and \mbh. For this purpose, a library of the accretion disk spectrum template was generated for a large range of [\ld, \mbh] pairs, e.g., [10$^{45}$ \lum, 10$^7$ \Msun], [10$^{45.1}$ \lum, 10$^7$ \Msun], ..., and [10$^{50}$ \lum, 10$^{10}$ \Msun]. By comparing the accretion disk template for a given [\ld, \mbh] pair with the observed data points, considering uncertainties, the $\chi^2$ value was calculated. This exercise was repeated for all [\ld, \mbh] pairs, thus effectively creating a $\chi^2$ grid. A cubic spline function was fitted on the $\chi^2$ grid to determine the global minimum and 1$\sigma$, 2$\sigma$, and 3$\sigma$ confidence levels. The results of this analysis are shown in Figure~\ref{fig:1a}. The best-fit log scale \mbh~and \ld~estimated were 9.81$^{+0.19}_{-0.20}$ (in \Msun) and 46.86$^{+0.09}_{-0.09}$ (in \lum), respectively. Within uncertainties, these parameters are similar to those obtained from the optical spectroscopic analysis.

\subsection{Comparison with Fermi-LAT and Swift-BAT Blazars}
The left panel of Figure~\ref{fig:2} shows the \mbh~distribution of all Fermi-LAT detected broad emission line blazars studied in \citet[][]{2021ApJS..253...46P}. For comparison, the \mbh~distribution of the \gm-ray emitting sources, also detected with the Swift-BAT, is overplotted. The average \mbh~(in \Msun) of blazars detected with BAT and LAT is $\log~\langle M_{\rm BH}\rangle=8.95$ which is higher than $\log~\langle M_{\rm BH}\rangle=8.52$ estimated for LAT-only detected blazars, though the scatter in the distributions is large, $\sim$0.6 dex. This result is aligned with the hypothesis that Swift-BAT has detected the luminous members of the blazar population hosting massive black holes. The examination of the location of PMN~J1310$-$5552 in this diagram has revealed it to be among the sources hosting some of the most massive black holes.

The distributions of \ld~for BAT-LAT and LAT-only detected broad line blazars studied in \citet[][]{2021ApJS..253...46P} are shown in the right panel of Figure~\ref{fig:2}. This plot highlights that the Swift-BAT detected blazars host more luminous accretion disks ($\log~\langle L_{\rm disk}\rangle=46.22$, in \lum) than LAT-only blazars ($\log~\langle L_{\rm disk}\rangle=45.73$, in \lum), though the scatter is large, $\sim$0.8 dex. Considering PMN~J1310$-$5552, its accretion power is similar to blazars hosting the most luminous accretion disks.

Overall, it can be concluded that PMN~J1310$-$5552 harbors one of the most powerful central engines among all \gm-ray detected blazars.

\begin{figure*}
\begin{center}
\hbox{
\includegraphics[scale=0.5]{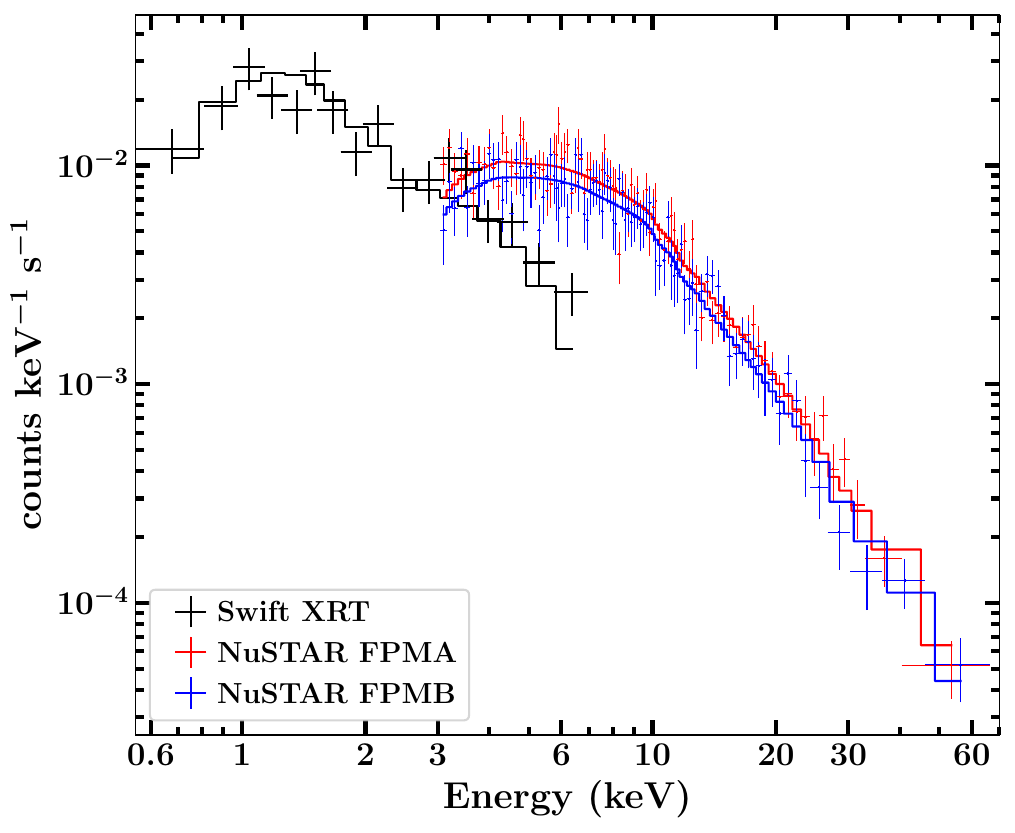}
\includegraphics[scale=0.55]{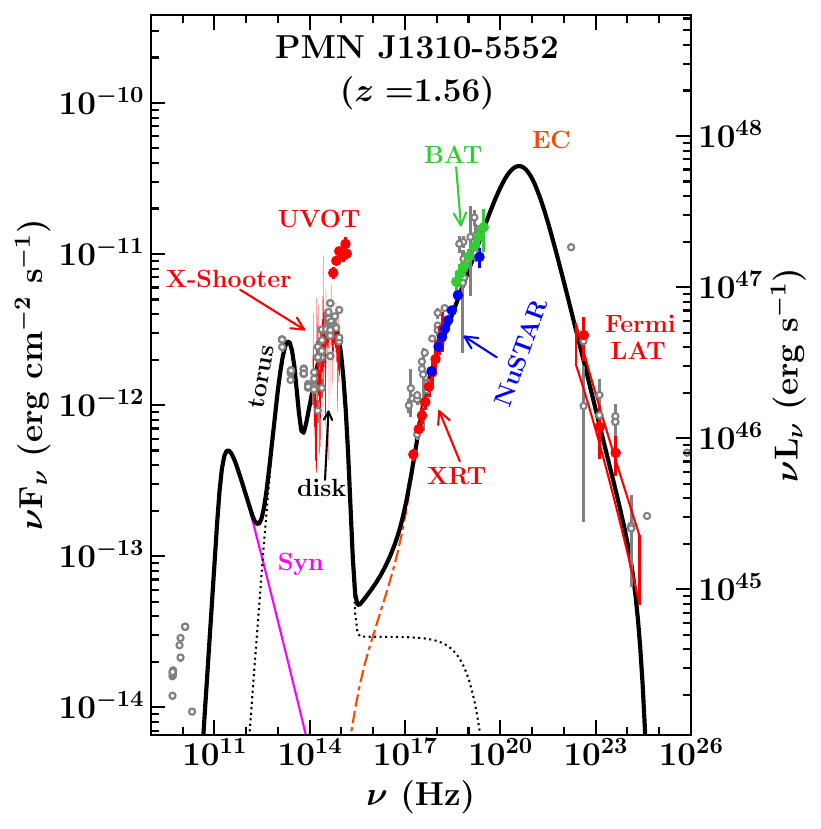}
}
\caption{Left: The data and the folded absorbed power-law model for the joint Swift-XRT and NuSTAR spectral fitting carried out in the 0.3$-$79 keV energy range. The spectra were binned to have at least 3$\sigma$ significance in each bin. Right: The broadband SED of PMN~J1310$-$5552. Various radiative components are shown with different lines as labeled, and the black solid line is the sum of all of them. Grey data points refer to archival observations adopted from the SSDC server.} \label{fig:3}
\end{center}
\end{figure*}

\section{Broadband Properties}\label{sec4}
\subsection{Classification}
PMN~J1310$-$5552 lacks blazar classification in the 4FGL-DR4 catalog. Its optical spectrum consists of broad emission lines (Figure~\ref{fig:1}), thus classifying it as an FSRQ. Its accretion luminosity in Eddington units is 0.1, i.e., 10\% of $L_{\rm Edd}$, which indicates a radiatively efficient accretion, further supporting the FSRQ classification \citep[][]{2011MNRAS.414.2674G}.

\subsection{Spectral Energy Distribution: The Data}
The broadband SED of a blazar provides crucial information about the physical properties of its relativistic jet and the central engine. Therefore, data from several observing facilities were collected and analyzed, as briefly described below.

{\it Fermi-LAT}: PMN~J1310$-$5552 is a faint \gm-ray emitter and is detected at $\sim$6$\sigma$ confidence level considering the first 14 years of the Fermi-LAT operation \citep[][]{2020ApJS..247...33A,2023arXiv230712546B}. Its \gm-ray spectral data points provided in the 4FGL-DR4 catalog were adopted for the broadband SED generation.

{\it NuSTAR}: Being a Swift-BAT detected object, PMN~J1310$-$5552 is included in the list of sources to be observed with NuSTAR as a part of its Extragalactic survey. The author requested the NuSTAR science operations center to prioritize observing this object, which was approved. PMN~J1310$-$5552 was observed on 2023 June 10 for a net exposure of 21.77 ksec (obsid: 60160528002). The data was analyzed following standard procedure\footnote{\url{https://heasarc.gsfc.nasa.gov/docs/nustar/analysis/nustar\_swguide.pdf}}. The Heasoft version 6.33.2 was used along with standard calibration files (version 20240311). The event file was cleaned and calibrated using the task {\tt nupipeline}, and the source and the background spectra were extracted using the pipeline {\tt nuproducts}. The source region was chosen as a circle of 49$^{\prime\prime}$ centered at the target quasar. The background region was considered as a circle of 70$^{\prime\prime}$ from the same chip but free from source contamination. The spectrum was binned to have at least 15 counts per bin, and the spectral fitting was performed in XSPEC \citep[][]{1996ASPC..101...17A}. The source was above the background up to $\sim$50 keV. A power-law model provided a reasonable fit ($\chi^2=80.5$ for 95 degrees of freedom). The estimated 3$-$79 keV energy flux is $1.88^{+0.15}_{-0.15}\times10^{-11}$ \ergflux~and photon index is $1.44^{+0.06}_{-0.07}$.

{\it Swift}: Simultaneous to the NuSTAR pointing, PMN~J1310$-$5552 was observed by the Swift satellite on 2023 June 10 for a net exposure of 6.2 ksec (obsid: 81135003). The X-Ray Telescope (XRT) data was analyzed using the online Swift-XRT data products generator\footnote{\url{https://www.swift.ac.uk/user\_objects/}} \citep[][]{2009MNRAS.397.1177E}. The spectrum was rebinned to have at least 15 counts per bin, and the spectral modeling was done in XSPEC adopting an absorbed power-law model ($\chi^2=13.3$ for 15 degrees of freedom). The Galactic neutral hydrogen column density along the line of sight ($N_{\rm H}=2.08\times10^{21}$ cm$^{-2}$) was taken from \citet[][]{2005AA...440..775K} and kept frozen during the fit. The unabsorbed energy flux and photon index in the 0.3$-$10 keV energy range were found to be $5.16^{+0.76}_{-0.59}\times10^{-12}$ \ergflux~and $1.12^{+0.18}_{-0.17}$, respectively.

The joint fitting of the NuSTAR and XRT spectra was also carried out by considering an absorbed power-law model with $N_{\rm H}$ fixed to the Galactic value (Figure~\ref{fig:3}, left panel). To take into account the cross-calibration uncertainties between different instruments, an intercalibration constant was inserted during the fit. This parameter was allowed to vary for the Swift-XRT and NuSTAR focal plane module A (FPMA) and frozen to unity for FPMB. It was estimated to be $0.83\pm0.10$ and $1.09\pm0.05$ for XRT and FPMA, respectively. In the energy range of 0.3$-$79 keV, the best-fitted photon index and normalization values were estimated to be $1.40^{+0.05}_{-0.05}$ and $5.74^{+0.67}_{-0.61}\times10^{-4}$ ph cm$^{-2}$ s$^{-2-1}$ keV$^{-1}$ ($\chi^2=173.27$ for 207 degrees of freedom). In order to examine the presence of any spectral break in the X-ray spectrum of PMN J1310$-$5552, the fitting was also carried out by applying an absorbed broken power law model. However, the fitting did not significantly improve with respect to the absorbed power-law model (F-test null hypothesis probability=0.04).

PMN~J1310$-$5552 was observed by the UltraViolet and Optical Telescope (UVOT) in all six filters. The data was analyzed following the recommended methodology. In particular, the individual frames were combined using the tool {\tt uvotimsum}, and the source magnitudes were extracted using the task {\tt uvotsource}. The source region was selected as a circle of radius 3$^{\prime\prime}$ centered at the blazar. The background, on the other hand, was a circle of 10$^{\prime\prime}$ radius from a nearby source-free region. The observed magnitudes were corrected for Galactic extinction and converted to energy flux units using the recommended conversion factors \citep[][]{2011AIPC.1358..373B}.

{\it Archival Observations}: The archival spectral measurements of PMN~J1310$-$5552 were collected from the Space Science Data Center\footnote{\url{https://tools.ssdc.asi.it/SED/}} (SSDC). Additionally, its 105-month averaged Swift-BAT spectrum was also considered. The spectral data points were extracted by fitting a simple power law model.

\subsection{Spectral Energy Distribution: The Model}
The broadband SED of PMN~J1310$-$5552 was modeled with the commonly adopted single-zone leptonic radiative model \citep[e.g.,][]{2009MNRAS.397..985G}. The spherical emission region was assumed to move along the jet with the bulk Lorentz factor $\Gamma$ and cover the whole cross-section of the jet. The emission region was considered to be filled with relativistic electrons whose energy distribution follows a smooth, broken power-law spectral shape. In a uniform and tangled magnetic field, these particles radiate via synchrotron, synchrotron self-Compton (SSC), and external Compton (EC) processes. The adopted reservoirs of the seed photons for the EC mechanism are BLR, dusty torus, and the accretion disk. The BLR and torus were assumed to be spherical shells of radii 
\begin{equation}
    R_{\rm BLR}=10^{17}\sqrt{\frac{L_{\rm disk}}{10^{45}}}~{\rm cm}; R_{\rm DT}=2.5\times10^{18}\sqrt{\frac{L_{\rm disk}}{10^{45}}}~{\rm cm},
\end{equation}
respectively, emitting blackbody radiation. The radiative energy density profiles of these components were adopted from \citet[][]{2009MNRAS.397..985G}. The BLR and torus were considered to reprocess 10\% and 30\% of the accretion disk radiation, respectively. The jet powers were derived following \citet[][]{2008MNRAS.385..283C}.

The SED model used in this work does not employ any statistical fitting. The accuracy of the adopted/derived physical parameters depends on the availability of contemporaneous multiwavelength observations. For example, the information on the central engine parameters, i.e., \mbh~and \ld, constrains the radiative profile and size of several external photon fields. In powerful blazars, both X- and \gm-ray spectra are well explained by the EC mechanism with negligible contribution from the SSC process in X rays, which regulates the magnetic field and size of the emission region \citep[see, e.g.,][]{2015ApJ...804...74P,2016ApJ...825...74P}. Furthermore, the shape of the X- and \gm-ray spectra directly constrains the spectral slopes of the broken power-law electron energy distribution. The X- and \gm-ray spectral shapes also enable an accurate measurement of the high-energy peak location, which was found to lie at $\sim$1 MeV. Accordingly, the synchrotron emission peaks at low frequencies ($<$10$^{12}$ Hz), leaving the accretion disk emission visible. Unlike the high-energy emission constraining the EC process, the lack of the multi-frequency radio observations of PMN~J1310$-$5552 implies that the synchrotron peak flux level was not well constrained, also keeping in mind the synchrotron self-absorption. In such a situation, the modeling was done considering the SED parameters obtained for large samples of `MeV' blazars in previous works \citep[e.g.,][]{2020ApJ...889..164M,2020ApJ...897..177P}. For example, the average Compton dominance for FSRQs cannot be very large, i.e., $\gtrsim$100 \citep[][]{2021ApJS..253...46P}. Since the high-energy SED peak is well-constrained, the synchrotron peak cannot be very low. If the synchrotron flux level is low, to explain the bright, high-energy radiation, one needs a large bulk Lorentz factor ($>$20$-$30), which is unlikely \citep[see, e.g.,][]{2014Natur.515..376G,2017ApJ...851...33P}. Further SED modeling guidelines can be found in \citet[][]{2017ApJ...851...33P}.

\subsection{Physical Properties}
The modeled SED covering radio-to-\gm-ray observations of PMN~J1310$-$5552 is shown in Figure~\ref{fig:3} (right panel), and associated parameters are provided in Table~\ref{tab:sed}.

\begin{table}
\caption{Summary of the SED Modeling Parameters.}\label{tab:sed}
\begin{tabular}{ll}
\tableline
Redshift                                             & 1.56     \\
Black hole mass, in \Msun, in log scale              & 9.81     \\
Accretion disk luminosity, in \lum, in log scale     & 46.86    \\
%Temperature of the dusty torus, in Kelvin            & 500      \\
Jet viewing angle, in degrees                        & 3$^{\circ}$ \\
Slope of particle distribution before break energy   & 1.7      \\
Slope of particle distribution after break energy    & 4.7      \\
Minimum Lorentz factor                               & 1        \\
Break Lorentz factor                                 & 49      \\
Maximum Lorentz factor                               & 3000     \\
Magnetic field, in Gauss                             & 1.2      \\
Bulk Lorentz factor                                  & 10       \\
Dissipation distance, in parsec                      & 0.25     \\
Size of the BLR, in parsec                           & 0.27     \\
\hline
Electron jet power, in \lum, in log scale        & 45.1   \\
Magnetic Jet power, in \lum, in log scale        & 45.8   \\
Radiative jet power, in \lum, in log scale           & 46.9   \\
Jet power in protons, in \lum, in log scale          & 47.6   \\
\tableline
\end{tabular}
\end{table}

\begin{figure*}
\hbox{\hspace{1.cm}
\includegraphics[scale=0.5]{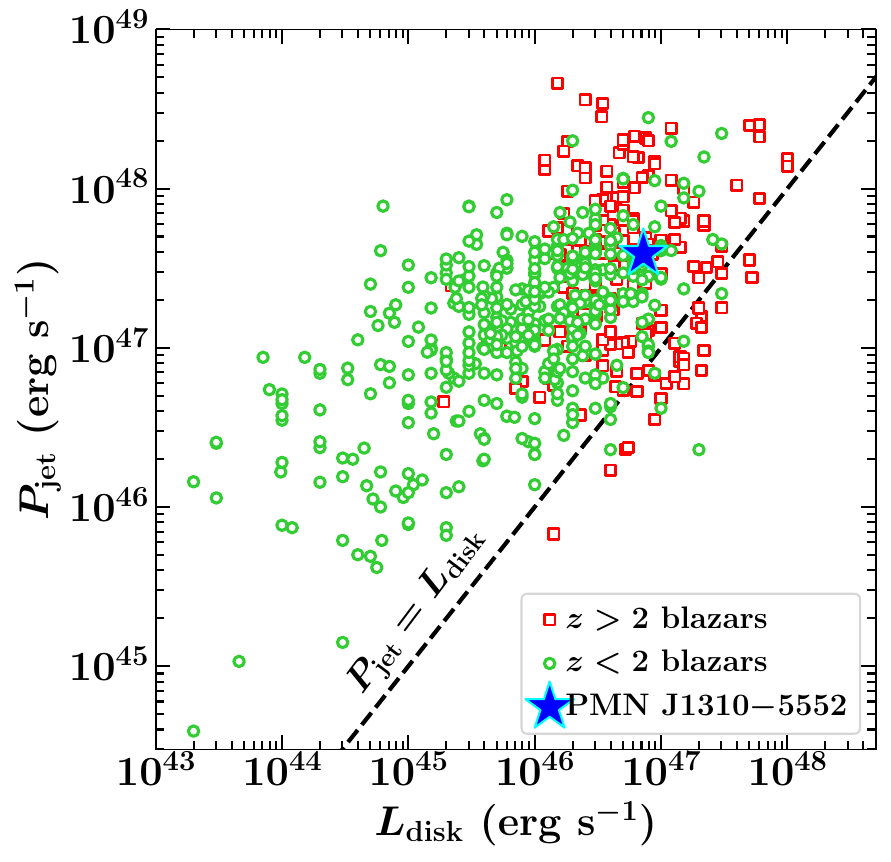}
\includegraphics[scale=0.5]{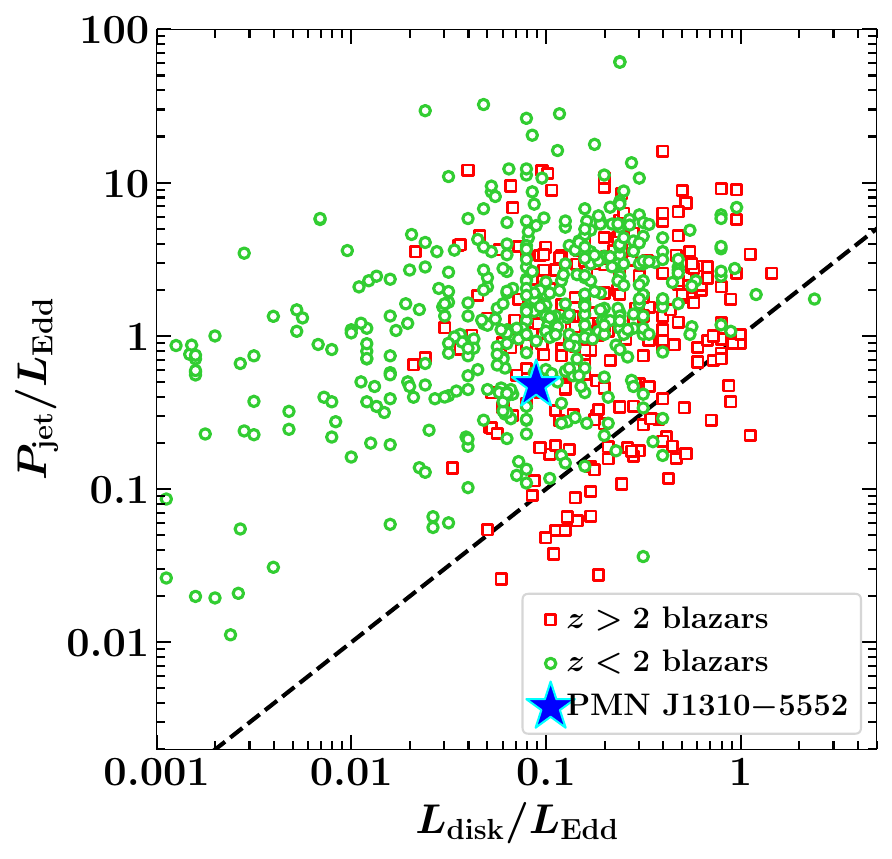}
}
\caption{The plot of the accretion disk luminosity versus jet power. The right plot shows the same quantities normalized for the mass of the black hole. The black dashed line represents the one-to-one correlation.} \label{fig:5}
\end{figure*}

The simultaneous NuSTAR and XRT data appeared consistent with the archival observations, thus suggesting that the source was not in any elevated activity state on 2023 June 10. The mission-averaged Fermi-LAT and Swift-BAT spectra also represented the typical activity of PMN~J1310$-$5552. Interestingly, the Swift-UVOT spectral data points are far from the archival optical observations, including the X-Shooter spectrum (Figure~\ref{fig:3}). Given the low-synchrotron peaked nature of the source, the rising spectral shape of the UVOT data does not match with the rest of the SED. This ruled out the possibility of flaring jet activity contributing to the elevated UVOT flux level. Previous Swift observations of PMN~J1310$-$5552 taken in 2006 December were also analyzed to examine the possible flux enhancement on 2023 June 10; however, no significant variability was identified. If one considers the high flux level of UVOT data points on two epochs (2006 December and 2023 June) due to elevated accretion disk activity, such a long-lasting accretion disk flaring episode is unlikely to occur. Also, even if the accretion disk emission remained at an elevated flux level for such a long time, it should have been reflected in other archival observations taken in this period, including X-Shooter and Goodman spectra. A possible explanation for the observed discrepancy could be that this object lies in the Galactic plane where the number density of stars is high. There might be contamination from UV bright stars lying adjacent to PMN~J1310$-$5552 to the observed source flux. Indeed, the earlier optical spectrum taken by \citet[][]{2008A&A...482..113M} turned out to be possibly that of a star so that the source confusion might be responsible for the observed brighter UVOT spectral data points.

The infrared-to-UV emission of PMN~J1310$-$5552 is well explained by the dusty torus and accretion disk models (Figure~\ref{fig:3}), which allowed to constrain the \mbh~and \ld~as discussed in Section~\ref{subsec:bhmass}. The source exhibits a flat, rising X-ray spectrum, and the \gm-ray emission has a steep falling spectral shape. Given the large bolometric power of the source, these observations are consistent with the reported anti-correlation of the SED peak frequencies and non-thermal jet luminosity \citep[e.g.,][]{2010ApJ...710...24S}. A flat X-ray spectrum also suggests the dominance of the EC over SSC since the latter is expected to have a soft spectral shape in the X-ray band due to the low-synchrotron peak nature of PMN~J1310$-$5552. The observed X-ray spectrum also constrained the low-energy cutoff of the electron population (\gm$_{\rm min}$), which has a rather low value for `MeV' blazars \citep[cf.][]{2017ApJ...839...96M}. The whole X- to \gm-ray spectrum was well reproduced by the inverse Compton scattering of the BLR photons. This sets the location of the emission region within the BLR, similar to other luminous blazars \citep[e.g.,][]{2020ApJ...897..177P}. Other SED parameters, e.g., magnetic field and bulk Lorentz factor, were also found to be similar to powerful FSRQs \citep[see, e.g.,][]{2017ApJ...839...96M,2019ApJ...871..211P}.

It is now well established that the jet power correlates with the accretion luminosity \citep[][]{1991Natur.349..138R,2014Natur.515..376G}. The location of PMN~J1310$-$5552 in the disk-jet power correlation plot is shown in Figure~\ref{fig:5}. The diagrams also include other blazars studied in previous works \citep[][]{2017ApJ...851...33P,2019ApJ...881..154P,2020ApJ...897..177P}. As seen in the left panel of Figure~\ref{fig:5}, PMN~J1310$-$5552 hosts one of the most powerful jets and accretion disks among all blazars below $z=2$. It has a jet power larger than \ld~similar to most other sources. The right panel of Figure~\ref{fig:5} shows the disk-jet power correlation normalized for \mbh, i.e., in Eddington units. For PMN~J1310$-$5552, the jet power and \ld~are found to be sub-Eddington. Some objects exhibit super-Eddington jet power; however, it could be due to the underlying assumption of the equal number density of electrons and protons, i.e., no electron-positron pairs. The budget of the jet power will reduce if the jet contains a few pairs \citep[see, e.g.,][]{2000ApJ...534..109S,2016ApJ...831..142M,2017MNRAS.465.3506P}.

\section{Summary}\label{sec5}
To identify luminous blazars harboring monstrous black holes among Fermi-BCUs, the author examined the 4FGL-DR4 and the BASS-DR2 catalogs. Only one source, PMN~J1310$-$5552, was identified with $M_{\rm BH}=1.05\times10^{10}$ \Msun~\citep[][]{2022ApJS..261....5M}. The detection of broad optical emission lines and radiatively efficient accretion classify it as a powerful FSRQ. The log scale central engine parameters, i.e., \mbh~and \ld, estimated from the analysis of the Goodman optical spectrum are 9.90$\pm$0.07 (in \Msun) and 46.86$\pm$0.03 (\lum), respectively. The modeling of the observed big blue bump with the standard accretion disk model resulted in the log scale  \mbh$=9.81^{+0.19}_{-0.20}$ (in \Msun) and \ld$=46.86^{+0.09}_{-0.09}$ (in \lum), respectively. The \mbh~and \ld values estimated from two different techniques are compatible and suggest PMN~J1310$-$5552 to host one of the most massive black holes and the most luminous accretion disks among the blazar population. The physical properties of this enigmatic blazar were explored by modeling the broadband SED using data from NuSTAR, Swift, Fermi-LAT, and archival observations. PMN~J1310$-$5552 is a `MeV' blazar with SED parameters similar to other members of this unique class of blazars.

The reported findings highlight that cosmic treasures are still hidden among the unknown high-energy emitters. Effective utilization of the ongoing wide-field, multiwavelength surveys and optical spectroscopic followup are the keys to discovering them and enhancing our understanding of the physics of relativistic jets.

\acknowledgements
The author thanks the journal referee for constructive criticism. Thanks are also due to the NuSTAR Science Operations Center for approving the request to prioritize observing PMN~J1310$-$5552. Thanks to Swift Satellite's principal investigator, Brad Cenko, for approving the observation request simultaneously with the NuSTAR pointing. Part of this work is based on archival data, software, or online services provided by the Space Science Data Center - ASI. This research has made use of NASA’s Astrophysics Data System Bibliographic Services. This research has made use of data obtained through the High Energy Astrophysics Science Archive Research Center Online Service, provided by the NASA/Goddard Space Flight Center. This work has made use of data from the NuSTAR mission, a project led by the California Institute of Technology, managed by the Jet Propulsion Laboratory, and funded by the National Aeronautics and Space Administration. This research has made use of the NuSTAR Data Analysis Software (NuSTARDAS) jointly developed by the ASI Science Data Center (ASDC, Italy) and the California Institute of Technology (USA).

\facilities{Swift, NuSTAR}

\software{XSPEC \citep[v 12.10.1;][]{1996ASPC..101...17A}, Swift-XRT data product generator \citep[][]{2009MNRAS.397.1177E}}

\bibliographystyle{aasjournal}
\bibliography{Master}

\end{document}